\documentclass[%
 reprint,
 amsmath,amssymb,
 aps,
]{revtex4-2}

\usepackage{graphicx}% Include figure files
\usepackage{dcolumn}% Align table columns on decimal point
\usepackage{bm}% bold math
\usepackage{orcidlink}
\usepackage{subfigure}
\usepackage{multirow}
\usepackage{bbold}
\usepackage{xcolor}
\usepackage{ulem}

\begin{document}

\preprint{APS/123-QED}

\title{Multi-wavelength analysis of the progenitor of GRB 230307A via Bayesian model comparison}% Force line breaks with \\

\author{Viviane Alfradique}
\email{vivianeapa@cbpf.com}
 \affiliation{
 Centro Brasileiro de Pesquisas F\'isicas, Rua Dr. Xavier Sigaud 150, \\
 22290-180 Rio de Janeiro, RJ, Brazil
}%
\author{Rodrigo da Mata}%
\affiliation{%
 Centro Brasileiro de Pesquisas F\'isicas, Rua Dr. Xavier Sigaud 150, \\
 22290-180 Rio de Janeiro, RJ, Brazil
}%
\author{Juan C. Rodr\'iguez-Ram\'irez}
\affiliation{Centro Brasileiro de Pesquisas F\'isicas, Rua Dr. Xavier Sigaud 150, \\
22290-180 Rio de Janeiro, RJ, Brazil
}%

\author{Clécio R. Bom}%
\affiliation{%
 Centro Brasileiro de Pesquisas F\'isicas, Rua Dr. Xavier Sigaud 150, \\
 22290-180 Rio de Janeiro, RJ, Brazil
}

\date{\today}% It is always \today, today,

\begin{abstract}
GRB 230307A is one of the brightest long-duration gamma-ray bursts (GRBs) ever detected, yet its progenitor remains uncertain due to the variety of plausible astrophysical scenarios. In this work, we investigate four possible progenitors for GRB 230307A: a binary neutron star (BNS), a neutron star–white dwarf (NS–WD) system, a neutron star–black hole (NS–BH) merger, and a tidal disruption event (TDE) involving a white dwarf and a supermassive black hole. Additionally, we explore three distinct central engine models powering the kilonova associated with the BNS: radioactive decay of $r$-process nuclei in a two-component ejecta model, a magnetar-driven model including magnetic dipole spin-down, and a combined model of magnetar spin-down with ${}^{56}\mathrm{Ni}$ radioactive decay. We perform Bayesian multi-wavelength light-curve analyses using physically motivated models and priors, and evaluate model performance through Bayes factors and leave-one-out cross-validation (LOO) scores. Our results show a statistical preference for a BNS or NS–WD progenitor producing a kilonova powered by a magnetar and ${}^{56}\mathrm{Ni}$ decay, characterized by a ${}^{56}\mathrm{Ni}$ mass of $\sim 4\times10^{-4}\,\mathrm{M}_{\odot}$ and an ejecta mass of $0.06\,\mathrm{M}_{\odot}$. Furthermore, under the assumption of a BNS origin within this model, we infer binary component masses of $m_{1} = 1.81^{+0.46}_{-0.61}\,\rm{M}_{\odot}$ and $m_{2} = 1.61^{+0.65}_{-0.41}\,\rm{M}_{\odot}$, with a dimensionless tidal deformability of $\tilde{\Lambda} = 471^{+318}_{-395}$. From the component mass posteriors, we infer that the observed offset can be explained by a natal kick as long as the systemic velocity is nearly aligned with the pre-kick orbital motion. In this case, the required kick velocity (co-moving frame) and binary separation range within $v'_\mathrm{k}\sim100$–$150~\mathrm{km\,s^{-1}}$, and $a_0\sim2$–$3~R_\odot$, respectively.
\end{abstract}
%Use showkeys class option if keyword display desired

\maketitle

The discovery of GRB 230307A, reported on March 7, 2023, by NASA's Fermi Gamma-ray Space Telescope \cite{Fermi2023}, Gravitational Wave High-energy Electromagnetic Counterpart All-sky Monitor (GECAM, \cite{GECAM2023}) and the Konus-Wind \cite{Konus2023}, represents one of the most luminous gamma-ray bursts ever observed. This event exhibited a peak flux of \(4.48 \times 10^{-4} \, \text{erg} \, \text{cm}^{-2} \, \text{s}^{-1}\) and a high gamma-ray fluence of \(3 \times 10^{-3} \, \text{erg} \, \text{cm}^{-2}\) within the 10--1000 keV energy range \cite{Sun2025}. The burst had a total duration of 42 seconds \cite{Sun2025}, and even 61 days after the event \cite{Levan2024, Yang2024}, a faint counterpart remained detectable across multiple wavelengths. Follow-up observations spanned the optical, near-infrared, and soft X-ray bands, with data collected by instruments such as the James Webb Space Telescope (JWST, \cite{Levan2024}), Hubble Space Telescope (HST, \cite{Yang2024}), Very Large Telescope (VLT, \cite{Levan2024}), Gemini South Telescope \cite{Levan2024}, Swift X-ray Telescope (XRT, \cite{Levan2024}), Chandra X-ray Observatory \cite{Levan2024}, Australia Telescope Compact Array (ATCA, \cite{Levan2024}), MeerKAT \cite{Levan2024}, AGILE \cite{Casentini2023}, and the Lobster Eye Imager for Astronomy (LEIA, \cite{Sun2025}). These comprehensive observations enabled a detailed characterization of the burst's afterglow and environment. The rapid decay of the bolometric luminosity, along with the increase in the photosphere radius, suggests the presence of a thermal component that associates a kilonova with the GRB and supports the presence of lanthanides in the ejected material \cite{Levan2024, Yang2024}. 

One way to investigate the origin of the progenitor is to analyze how variations in kilonova evolution are intrinsically related to the properties of the ejecta and the remnant of the merger. This analysis provides valuable insights into the physical processes that are consistent with observations while excluding those that are not supported by the data. In the case of GRB 230307A, the rapid decay of the optical emission at early times, followed by the dominance of emission in the near-infrared (NIR) band, may indicate the presence of heavy isotopes \cite{Yang2024, Levan2024}. Furthermore, the detection of tellurium in the mid-infrared spectrum provides additional support for $r$-process nucleosynthesis, which is expected in BNS and NS-BH mergers. These observations, together with the significant offset between the burst position and its potential host galaxy at redshift $z=0.065$, suggest a compact binary merger origin, possibly explained by a high kick velocity imparted to the neutron star component \cite{Kalogera1998}.

Another notable feature of GRB 230307A is the soft X-ray emission observed by LEIA \cite{Sun2025} in the 0.5–4.0 keV band. This emission shows a plateau during the first 10 seconds, followed by a decay, a behavior commonly interpreted as the spin-down signature of a magnetar formed after the compact binary merger. However, the nature of the magnetar central engine remains debated. As noted in \cite{Wang2024}, neutrino emission could suppress the production of lanthanide-rich ejecta, and the mechanisms connecting spin-down luminosity to X-ray emission are still not fully understood \cite{Du2024}. In light of these challenges, Wang et al. \cite{Wang2024} proposed a NS–WD merger with a possible remnant magnetar as the progenitor. In this scenario, the long-lived emission is explained by the lower density of the white dwarf, with heavy elements produced by $^{56}\text{Ni}$ decay rather than the $r$-process typical of BNS mergers. They showed that an NS–WD system with a $\sim10^{-3}\,\rm{M}_{\odot}$ white dwarf and an ejecta mass of $\sim0.1\,\rm{M}_{\odot}$ provides a consistent fit to the multi-wavelength afterglow and kilonova emission. Beyond the magnetar engine, other models have also been proposed, such as disruption of the NS crust during the inspiral phase \cite{Tsang2012, Palenzuela2013, Neill2021, Suvorov2021} or magnetospheric interactions between the binary components \cite{McWilliams2011, Most2020, Beloborodov2021, Cooper2023}. \cite{Du2024} was the first study to investigate the progenitor of the exotic GRB 230307A. Their analysis relied on the phenomenological classification of GRBs into type I, associated with massive star collapses, and type II, resulting from compact object mergers, as proposed by \cite{Lu2010,Lu2014}. They found that GRB 230307A has an $\epsilon$ value (defined as the ratio between the isotropic gamma-ray energy and the rest-frame spectral peak energy) lying on the boundary between the two classes, with a $3\sigma$ deviation from the type II classification. However, the effective amplitude parameter they obtained is consistent with type II. Given the similarity of these values to those of other long GRBs, the authors concluded that GRB 230307A most likely originated from a compact object merger, either a BNS or an NS–WD system, although it remains unclear which of these channels is the progenitor.

Identifying the progenitor system responsible for transient astrophysical events, such as GRB~230307A, is a fundamental step in understanding the physics governing such explosions. The degeneracy in the observable signatures makes it crucial to develop robust model selection strategies. This is especially important as we enter the era of high-cadence, multi-messenger observations enabled by facilities such as the Vera C. Rubin Observatory, which will conduct the Legacy Survey of Space and Time \cite{abell2009lsst}, anticipated to reveal a large number of such events. A reliable identification of the progenitor will not only refine our models of compact binary evolution, but also improve our understanding of the equation of state (EOS) of dense matter, nucleosynthesis pathways, and the role of magnetic fields in shaping the observed electromagnetic emission.

In this work, we investigate multiple progenitor scenarios for GRB~230307A within a comprehensive Bayesian framework, aiming to identify the model that best reproduces the observed multiwavelength data, particularly the afterglow and kilonova components. We perform Bayesian inference using the nested sampling algorithm implemented in \textsc{dynesty} \cite{Speagle2020} and evaluate competing models with statistical tools including the Bayes factor, the LOO score, and the Kullback–Leibler (KL) divergence. These complementary metrics allow a robust model comparison by balancing goodness of fit, model complexity, and information content.

The paper is structured as follows: Section \ref{models} describes the observational data and theoretical models; Section \ref{methods} outlines the Bayesian methodology; Section \ref{results} presents the model comparison and parameter inference results; and Section \ref{conclusion} summarizes the implications for the identification of the GRB 230307A progenitor.

\section{Data and ejecta models}\label{models}
\subsection{Observational data}
The data used to perform the multi-wavelength analysis were collected from the source data provided by \cite{Yang2024} (cf. “Source Data Fig. 3”), which gathered data from the HST, JWST, Swift/XRT, Swift/UVOT, X-Shooter, Chandra, Gemini, XMM-Newton, and Fermi/GBM. The follow-up observations cover the optical with the Gemini telescope and the Southern Astrophysical Research telescope, near-infrared with the Gemini Telescope and the JWST, which also explored the mid-infrared and far-infrared bands; in the radio with the ATCA; and in the X-ray with Swift/XRT, Chandra X-ray Observatory, and XMM-Newton.

\subsection{Possible scenarios of GRB 230307A progenitor}\label{progenitormodel}
To investigate the electromagnetic counterpart of GRB 230307A, we adopt a modeling approach that separates the afterglow and kilonova components, allowing a clear comparison between different scenarios for thermal emission. We use the Python package \textsc{afterglowpy} \cite{Ryan2020} to model the synchrotron radiation from a Gaussian structured jet (unless stated otherwise), keeping the afterglow fixed across all analyses to isolate the effects of changing the kilonova emission models. The afterglow model includes standard forward shock emission, with free parameters: isotropic-equivalent kinetic energy $E_0$, circumburst density $n_0$, the magnetic- and electron-energy fractions $\varepsilon_B$ and $\varepsilon_e$, respectively, electron power-law index $p$, jet core opening angle $\theta_c$, and electron participation fraction $\xi_N$. Each kilonova model then explores distinct assumptions about the geometry, composition, and energy sources of the ejecta, allowing us to assess their individual contributions to the observed emission without introducing degeneracies from varying afterglow geometry.

\subsubsection{Binary Neutron Stars Merger}
We adopt the mixed-component kilonova model described in \citet{Metzger:2019zeh}, where the total emission arises from the sum of two distinct ejecta components: a blue (lanthanide-poor) and a red (lanthanide-rich) component. This framework captures the multichannel nature of neutron star merger ejecta, with fast, polar outflows typically being neutron-rich and of low opacity ($\kappa \sim 0.5$--$1~\mathrm{cm}^2\mathrm{g}^{-1}$), producing an early, short-lived blue kilonova. In contrast, slower equatorial tidal ejecta are generally lanthanide-rich and highly opaque ($\kappa \sim 5$--$10~\mathrm{cm}^2\mathrm{g}^{-1}$), resulting in a longer-lasting red kilonova that peaks in the near-infrared. The model assumes that the transient is powered by the radioactive decay of \textit{r}-process nuclei synthesized in the neutron-rich ejecta. Each component of the ejecta is treated as an expanding black body with fixed opacity and velocity and is characterized by its own mass, opacity, and thermalization efficiency. Following GW170817, this two-component, one-dimensional modeling approach has become standard practice. In multicomponent implementations, the components evolve independently, and their emissions are summed to obtain the total light curve. Here, we modified the implementation of this model provided by the Python package \textsc{nmma} \cite{Pang:2022rzc}.

\subsubsection{Binary Neutron Stars with central engine}
We adopt the engine-powered kilonova model introduced by \cite{Sarin2022} and implemented in the Python package \textsc{redback} \cite{redback}, which extends traditional radioactive kilonova frameworks to include energy injection from a long-lived, rapidly rotating neutron star (a magnetar) formed as a result of the binary neutron star merger. The model accounts for spin-down of the remnant via both magnetic dipole radiation and gravitational wave emission, depending on the strength and configuration of the internal toroidal and external dipolar magnetic fields. The resulting energy budget, governed by the competition between these two channels, significantly impacts the luminosity and temporal evolution of both the kilonova and its afterglow. In this framework, the magnetar wind interacts with the expanding ejecta, increasing its kinetic and internal energy while also modifying the radiative efficiency through time-dependent gamma-ray leakage. The model solves a set of coupled differential equations tracking the evolution of the Lorentz factor, internal energy, and radius of the ejecta, while accounting for radiative losses and adiabatic expansion.

\subsubsection{Neutron Star–Black Hole Merger}
We adopt the neutron star–black hole kilonova framework developed by \cite{Anand:2020eyg}, which models the expected optical/IR emission following NS–BH mergers by combining radiative transfer simulations with gravitational wave parameter constraints. We use the \textsc{redback} implementation of this model. Kilonova emission is modeled using the radiative transfer code \textsc{POSSIS} \cite{Bulla2019}, with an updated grid of synthetic spectra tailored to NS–BH systems. The model incorporates a two-component ejecta structure: lanthanide-rich dynamical ejecta concentrated in the equatorial plane, and a more isotropic post-merger wind component with intermediate opacity.

\subsubsection{Tidal disruption event}
We utilize the tidal disruption event model implemented in the Modular Open Source Fitter for Transients (\textsc{mosfit} \cite{Mockler2019} and \textsc{redback}), which systematically fits TDEs using a physically motivated framework. The model is based on hydrodynamical simulations of polytropic stars disrupted by supermassive black holes (SMBHs), from which the fallback rate of stellar debris is derived and converted into a bolometric light curve assuming a constant radiative efficiency. The fallback rate $\dot{M}_{\rm fb}$ depends on the black hole mass, stellar mass, stellar structure (parameterized by a polytropic index), and the impact parameter of the disruption event. To account for potential time delays due to circularization and disk accretion, the model introduces a viscous delay timescale that acts as a low-pass filter on the fallback rate. The resultant accretion-powered luminosity is then reprocessed through a photospheric layer, which is assumed to emit as a blackbody.

\subsubsection{Neutron Star-White Dwarf Merger}\label{magnetar}
We also consider the semi-analytical model proposed in \cite{Wang2024}, which interprets the optical and infrared emission of GRB~230307A within the framework of an NS–WD merger. In this scenario, the late-time kilonova-like emission is powered by both the spin-down energy of a long-lived magnetar and the radioactive decay of a small quantity of $^{56}$Ni, rather than by $r$-process nucleosynthesis, which is unlikely to occur in the low-density ejecta typical of NS–WD mergers. While the model is primarily applied to NS–WD mergers, it can also be extended to BNS mergers, as such systems are likewise expected to produce a magnetar accompanied by $^{56}$Ni synthesis \cite{Jacobi2025, Ai2025}. The model describes the merger ejecta as multiple concentric shells with fixed velocities and a power-law density profile. The spin-down luminosity of the magnetar follows a magnetic dipole formula with a fixed timescale derived from early X-ray data, while the $^{56}$Ni and $^{57}$Co radioactive heating is modeled using standard exponential decay laws. The energy evolution of each ejecta shell includes contributions from magnetar injection, radioactive decay, adiabatic expansion, and photon diffusion. The light curve is computed by summing the contributions from all shells and assuming blackbody emission with temperature given by the Stefan–Boltzmann law.

\section{Bayesian analysis tests}\label{methods}

\subsection{Bayes factor}
A straightforward way to compare two models in terms of ``goodness of fit",  based on observed data, is by evaluating the \textit{Bayes' ratio}. Let \textit{M} denote a model characterized by a set of parameters $\theta_{i}^{M}$. The Bayes' ratio is defined as the ratio of the evidence term $\mathcal{Z}$ (i.e., the integral of the product between the likelihood $\mathcal{L}$ and the prior $p\left(\theta_{i}^{M}\right)$ over the entire parameter space), 
\begin{equation}
    \mathcal{Z}^{M} = \int \mathcal{L}\left(d|\theta_{i}^{M}\right)p\left(\theta_{i}^{M}\right)d^{n}\theta_{i}^{M},
\end{equation}
of the two competing models:
\begin{equation}
   \ln B_{12} = \ln\left(\frac{\mathcal{Z}^{1}}{\mathcal{Z}^{2}}\right).
\end{equation}

The criterion for interpreting the Bayes factor is called Jeffreys' scale \cite{Jeffreys1961}. This scale classifies the strength of evidence into four categories in favor of model 1:
\begin{enumerate}
    \item Strong evidence: if $\ln B_{12}  > 5$;
    \item Moderate evidence: if $ 2 < \ln B_{12} \leq 5$;
    \item Weak evidence: if $0 < \ln B_{12} \leq 2$;
    \item No evidence (models are equally supported): if $\ln B_{12} = 0$;
\end{enumerate}
If $\ln B_{12}$ assumes negative values, then model 2 is better supported than model 1 and is classified as follows:
\begin{enumerate}
    \item Strong evidence: if $\ln B_{12} < -5$;
    \item Moderate evidence: if $ -5 \leq \ln B_{12} < -2$;
    \item Weak evidence: if $-2\leq \ln B_{12}  < 0$.
\end{enumerate}

Throughout this article, we will compute the Bayes factor relative to our best-ﬁtting model as a reference, where we will refer to as $\ln B_{1,\rm{ref}}\left(=\ln\left(\mathcal{Z}^{1}/\mathcal{Z}^{\rm{ref}}\right)\right)$.

\subsection{Leave-One-Out cross-validation}
Leave-One-Out cross-validation \cite{vehtari2017loo} is a fully Bayesian method for analyzing the predictive performance of a model by estimating how well it predicts each observation when that observation is left out of the fitting process. LOO evaluates the likelihood of each observation when left out of the fit, averaging over the full posterior distribution of the model parameters to approximate the expected log predictive density (elpd) for new data. The LOO estimate is defined as the sum of the logarithm of the posterior predictive probability of each left-out observation, $p(y_i \mid y_{-i})$:
\begin{equation}
\mathrm{elpd}_{\mathrm{LOO}} = \sum_{i=1}^{n} \log p(y_i \mid y_{-i}),
\end{equation}
where $y_i$ is the $i$-th observation and $y_{-i}$ denotes the data with the $i$-th observation removed. Higher $\mathrm{elpd}_{\mathrm{LOO}}$ values indicate better predictive performance. 

\subsection{Kullback-Leibler divergence}
The Kullback-Leibler divergence is a primary measure of the divergence between two probability distributions, $p\left(\vec{\theta}\right)$ and $q\left(\vec{\theta}\right)$, which is quantified in the amount of information lost when the distribution $q\left(\vec{\theta}\right)$ is used instead of the distribution $p\left(\vec{\theta}\right)$. The KL divergence, $D_{\rm KL}\left(p||q\right)$, is defined as
\begin{equation}
    D_{\rm KL}\left(p||q\right) = \int_{\vec{\theta}} p\left(\theta_{i}\right)\log\left(\frac{p\left(\theta_{i}\right)}{q\left(\theta_{i}\right)}\right)d^{n}\theta_{i},
\end{equation}
and quantifies how well $q\left(\vec{\theta}\right)$ approximates $p\left(\vec{\theta}\right)$. The primary Bayesian analysis in this work compares different kilonova models while fixing the afterglow model. In this way, our results explore the KL divergence for the posterior distribution obtained using the Markov Chain Monte Carlo method, focusing solely on the afterglow parameter space. The KL divergence values are interpreted such that values below 0.1 indicate negligible divergence, values between 0.1 and 1 suggest moderate divergence, and values greater than 1 denote substantial divergence between the posterior distributions.

\section{Results}\label{results}
In this section, we present our multi-wavelength analyses of GRB 230307A. In the first subsection, we describe our model selection analysis, aimed at investigating the nature of the progenitor and the ejecta mechanism that powered the observed kilonova emission. In subsection \ref{KNeffect}, we describe the influence of the kilonova emission on the GRB, and in subsection \ref{offset}, we explore the binary properties of the GRB 230307A progenitor and investigate the possibility of a supernova natal kick to explain the observed offset.

\subsection{Model selection: preferred progenitor system and central engine scenario}\label{modelselection}
\begin{figure*}
    \centering
    \includegraphics[width=\linewidth]{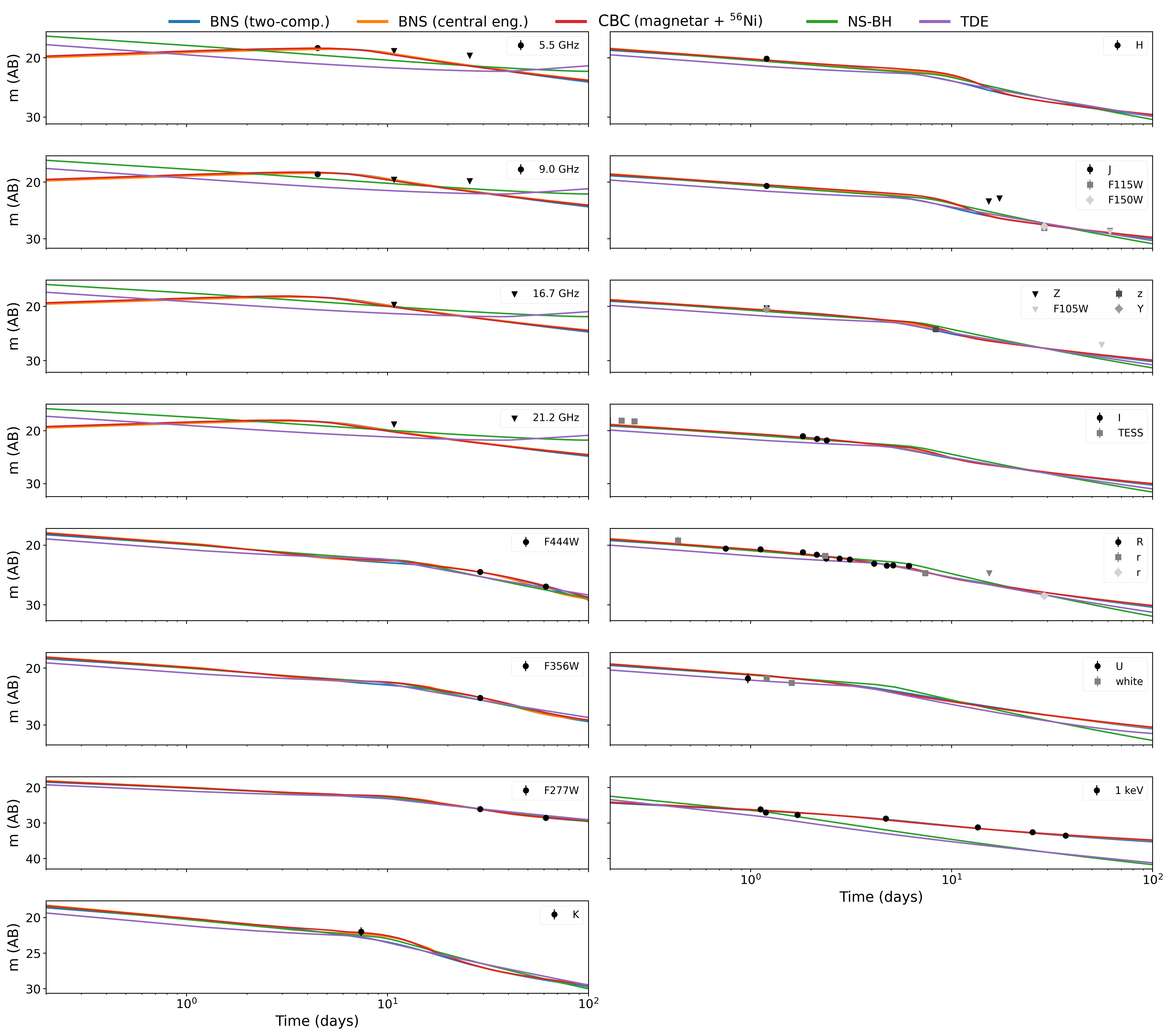}
    \caption{GRB 230307A multi-wavelength light curves. Solid lines show the best-fitting model curves obtained from joint Bayesian inference for different progenitor scenarios (shown in different colors; see App.\ref{app:inference_resu} for details). Black dots represent observational data points with uncertainties, and triangles denote upper limits. In some cases, the error bars are not visible because their size is smaller than the plotting scale.}
    \label{fig:lightcurve_all}
\end{figure*}

Here, we apply the statistical model selection methods described in Section \ref{methods} to a joint analysis of the GRB afterglow and a candidate kilonova signature or emission from a tidal disruption event. We investigate several compact merger progenitor scenarios that could produce the kilonova emission, including a compact binary (either a binary neutron star merger or a white dwarf–neutron star merger) and a neutron star–black hole merger. The median values and corresponding 1$\sigma$ uncertainties for the physical parameters of each model analyzed in this work are presented in Table \ref{tab:bestfit}. The results show reasonable agreement, within 1$\sigma$ to 2$\sigma$, with previous estimates reported in \cite{Yang2024} and \cite{Wang2024}. The parameter inference from the magnetar spin-down model indicates an ejecta mass of $0.06\,M_\odot$ and a $^{56}$Ni mass of $\sim 4\times 10^{-4}\,M_\odot$, consistent with that expected from a typical NS–WD merger \cite{Zenati2019}, while also remaining compatible with the higher, more characteristic $^{56}$Ni production of a BNS merger \cite{Jacobi2025}. Therefore, throughout the remainder of this paper, we adopt this model as a generic compact binary coalescence (CBC) magnetar spin-down scenario, without excluding either of these two progenitor possibilities. Table \ref{tab:events} summarizes our results for the different possible scenarios, presenting the Bayes factors, LOO, and maximum likelihood values. Based on the Bayes factor, we find that the scenario best supported by the data is the CBC magnetar spin-down model, which yields a maximum log-likelihood of –99.86. This corresponds to likelihood ratios of 1.1, 1.8, 39, and 85 relative to the BNS two-component kilonova, BNS general merger-nova, NS–BH, and TDE models, respectively. We adopt this model as the reference for the Bayes factor comparison. Figure \ref{fig:lightcurve_all} presents the multi-wavelength light curves of GRB 230307A for all models discussed in Section \ref{methods}, assuming the best-fit parameters and prior bounds listed in Table \ref{tab:bestfit}. 

We compare models using both Bayes factors and LOO scores (see results in Table \ref{tab:events}). The Bayes factor analysis provides strong evidence against all alternative models compared to the reference model, with logarithmic Bayes factor values below –5. The LOO cross-validation analysis indicates that the CBC magnetar spin-down model provides the best predictive accuracy, with $\mathrm{elpd}_{\mathrm{loo}} = -52 \pm 17$. The BNS two-component model exhibits lower predictive performance 
($\mathrm{elpd}_{\mathrm{loo}} = -141 \pm 42$), but its difference relative to the magnetar spin-down model corresponds to only $\sim 2\sigma$, which does not constitute decisive evidence against it. Therefore, while the CBC magnetar spin-down scenario is statistically preferred, the BNS two-component model remains a possible alternative. The BNS general merger-nova model yields an even lower $\mathrm{elpd}_{\mathrm{loo}}$, with a difference of approximately $2.6\sigma$ relative to the CBC magnetar spin-down model, indicating weaker support compared to the BNS two-component model, yet it cannot be entirely excluded. By contrast, the NS+BH and SBH+WD astrophysical scenarios have extremely low $\mathrm{elpd}_{\mathrm{loo}}$ values, suggesting poor predictive performance. However, the large standard errors imply that their true predictive ability is highly uncertain; thus, while these models are strongly disfavored, they cannot be strictly ruled out based solely on this analysis. These results reinforce the conclusions of \cite{Wang2024} and \cite{Yang2024}, providing strong evidence, in terms of both goodness of fit and model complexity, in favor of a scenario where the progenitor of GRB 230307A is the coalescence of a compact binary, compared to systems formed by NS–BH mergers producing kilonova emission, or SBH–WD mergers leading to tidal disruption events. 

To gain a deeper understanding of our results and establish connections between the underlying emission processes, we analyzed the best-fit parameters of each model in each wavelength band. The BNS two-component model provides the best fit to the NIR data, with a modest improvement in chi-squared ($\chi^2 \sim 3$) relative to the CBC magnetar spin-down model. In contrast, in the optical band, the CBC magnetar spin-down model performs better ($\chi^2 \sim 36$), while the BNS two-component ($\chi^2 \sim 45$) and BNS general merger-nova ($\chi^2 \sim 69$) models are slightly less favored. Both the NS-BH and TDE models are strongly disfavored in both bands, exhibiting significantly higher $\chi^2$ values ($\approx$ 160–1650), indicating a poor agreement with the observed thermal emission of GRB 230307A. The TDE model’s failure is likely due to its lack of lanthanide-rich ejecta, which are essential to reproduce the NIR flux. Additionally, the poor fit of the NS-BH model suggests that the observed data require a blue optical emission component, commonly attributed to lanthanide-poor disk winds, that is typically absent in standard NS-BH kilonova models. On the other hand, the radio and X-ray bands exhibit low $\chi^2$ values (approximately 0.5–4 for the CBC kilonova models and approximately 15–320 for the TDE and NS-BH models), reflecting the dominant contribution of the afterglow emission relative to the kilonova component in fitting the observed data. Our results show the best fit for the X-ray band with the CBC magnetar spin-down model, while the radio data are better described by the BNS two-component model. Overall, the CBC kilonova models provide a superior description of the afterglow-dominated data, as also reflected by their higher maximum likelihood values (see Table \ref{tab:events} for a comparison).

\begin{table*}
\centering
\caption{Summary statistical results for the multi-wavelength analysis of GRB 230307A.}
\begin{ruledtabular}
\begin{tabular}{ccccccc}
\multirow{2}{*}{Transient emission model} & \multirow{2}{*}{$\ln\left(\mathcal{Z}^{1}/\mathcal{Z}^{\rm{ref}}\right)$} & Maximum  
& \multirow{2}{*}{LOO} & \multirow{2}{*}{Dimensions} & KN model \\ 
 & & $\ln(\mathcal{L})$ & & & reference \\
\hline
BNS system & \multirow{2}{*}{-25.81} & \multirow{2}{*}{-110.69} 
& \multirow{2}{*}{-141$\pm$42} & \multirow{2}{*}{17} & \multirow{2}{*}{\cite{Villar2017}} \\
 (two-component kilonova) & & & & & \\
BNS system & \multirow{2}{*}{-87.08} & \multirow{2}{*}{-181.06} 
& \multirow{2}{*}{-52$\pm$17} & \multirow{2}{*}{13} & \multirow{2}{*}{\cite{Sarin2022}} \\
(General Merger-Nova) & & & & & \\
CBC system & \multirow{2}{*}{-} & \multirow{2}{*}{-99.86} 
& \multirow{2}{*}{-170$\pm$43} & \multirow{2}{*}{14} & \multirow{2}{*}{\cite{Wang2024}}\\
(magnetar spin-down) & & & & & \\
NS+BH system & \multirow{2}{*}{-3838.82} & \multirow{2}{*}{-3932.98} 
& \multirow{2}{*}{-3906$\pm$2854} & \multirow{2}{*}{15} & \multirow{2}{*}{\cite{Villar2017}} \\
(two-component kilonova - ejecta relation) & & & & & \\
SBH+WD system & -8388.02 & -8458.49 
& -5370$\pm$2946 & 15 & \cite{Mockler2019}\\
(TDE - 4/3 polytropes stars) & & & & & \\
\end{tabular}
\end{ruledtabular}
\label{tab:events}
\end{table*}

\subsection{Modeling the afterglow of GRB 230307A with and without kilonova emission}\label{KNeffect}

It has been established in previous studies that the emission from GRB 230307A is consistent with a gamma-ray burst accompanied by kilonova emission \cite{Yang2024, Wang2024}. In particular, these works showed that the inclusion of a kilonova component is essential to reproduce the enhanced brightness of the optical and NIR light curves compared to standard afterglow models. In this context, we analyze the ability of a GRB model with a Gaussian jet structure to fit the observed data and quantify the improvement achieved when including a kilonova component, described by the two best-fit models found in the previous subsection: the magnetar spin-down scenario and the two-component kilonova. Furthermore, we evaluate how the inclusion of transient emission affects the inferred parameters describing the afterglow.

\begin{figure}
    \centering
    \includegraphics[width=\linewidth]{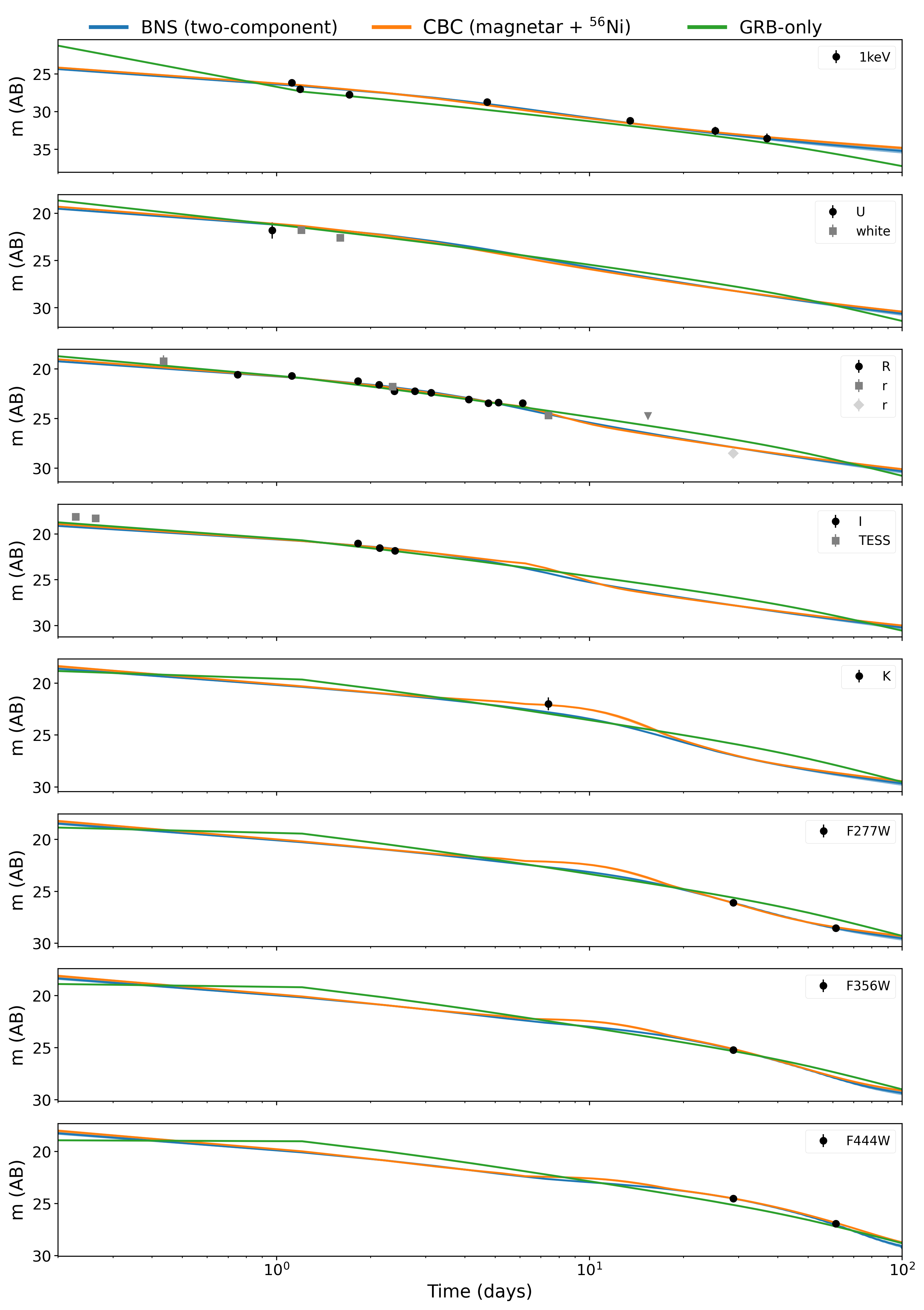}
    \caption{Multi-wavelength light curves of GRB 230307A from the afterglow-only emission (green) and with the addition of a kilonova, modeled using the BNS two-component model (blue) and the CBC magnetar spin-down model (orange), for the X-ray, optical, and NIR bands. Black dots represent observational data with uncertainties, and triangles indicate upper limits. For certain data points, the error bars are not shown because they are smaller than the scale of the plot.}
    \label{fig:lightcurve_grb}
\end{figure}

Our analysis shows that the afterglow-only model yields a maximum log-likelihood lower by factors of 23 and 25 compared to models in which the BNS two-component kilonova or CBC magnetar spin-down emission is added to the afterglow, respectively. This result is mainly governed by the NIR band, particularly the F444W and F277W filters, which provide well-measured data and yield considerable $\chi^2$ values of approximately 4030 and 564, respectively. The large $\chi^2$ values from these bands reveal the need for an additional emission component to better reproduce the observational data at late times (after $\approx 2$ days post-explosion; see the panel in Fig.~\ref{fig:lightcurve_grb}). The best-fitting light curve for the GRB-only model shows better agreement with the data in the X-ray and optical bands, yielding lower $\chi^2$ values in the range of 0.5–36.

\begin{figure*}
    \centering
    \begin{subfigure}{}
        \includegraphics[width=\linewidth]{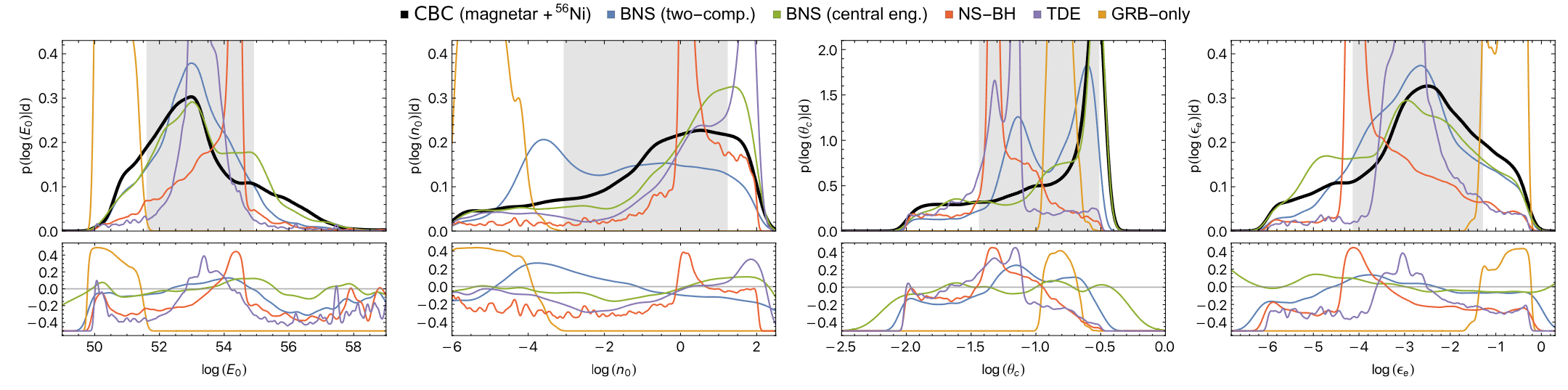}
    \end{subfigure}
    \begin{subfigure}{}
        \includegraphics[width=5.5in]{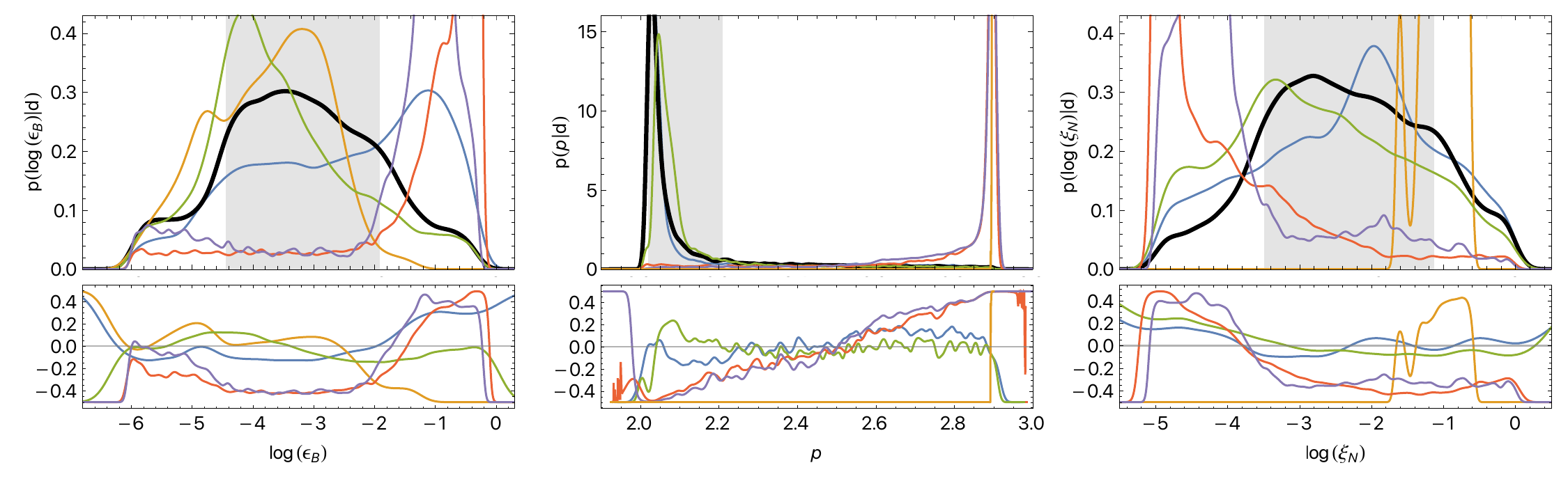}
    \end{subfigure}
    \caption{Posterior comparison of GRB afterglow model parameters obtained from joint analyses with different possible astrophysical scenarios. The subplots show the residuals relative to the CBC magnetar spin-down model, defined as $2\times\left(\rm{PDF_{KN\, model}}-\rm{PDF_{ref.}}\right)/\left(\rm{PDF_{KN\, model}}+\rm{PDF_{ref.}}\right)$. The shaded gray region denotes the 1$\sigma$ credible interval of the reference model's posterior distributions.}
    \label{fig:gaussianeff}
\end{figure*}

In Table \ref{tab:eventsKL}, we present the KL divergence values obtained for all transient emission models explored in this work, assuming the CBC magnetar spin-down model as the fiducial model. We find that most parameter distributions from the GRB-only inference, except for the magnetic energy fraction, diverge substantially from the fiducial model, with KL values exceeding 1. This divergence is further supported by the significant disagreement, greater than 1$\sigma$, between the posterior means (see Fig.~\ref{fig:gaussianeff}). The results obtained for the NS–BH and TDE models show partial agreement within 1$\sigma$, with moderate divergence from the true model for ${\log_{10} E_0, \log_{10} n_0, \log \epsilon_c, \log \theta_{c}}$, and significant divergence for the remaining parameters. In contrast, all BNS scenarios (two-component and general merger-nova models) exhibit negligible divergence for most parameters. This emphasizes that the inference of afterglow parameters is reliable only when the assumed progenitor scenario matches the true one. Otherwise, the inclusion of different transient components, or their omission, results in divergent Bayesian inferences and, consequently, different interpretations of the afterglow emission.

\begin{table}
\centering
\caption{KL divergence values computed from the posterior distributions of the Gaussian jet-structure parameter model across the different scenarios investigated in this work. The results obtained with the CBC magnetar spin-down model (see subsection \ref{magnetar}) are taken as the reference (true) probability distribution.}
\begin{ruledtabular}
\begin{tabular}{cccccccc}%{lllll}
  & \multirow{2}{*}{GRB-only} & BNS & BNS & \multirow{2}{*}{NS-BH} & \multirow{2}{*}{TDE}\\
  & & (two-comp.) & (central eng.) & & \\
 \hline
$\log_{10} E_0$ & 5.81 & 0.09 & 0.03 & 0.72 & 0.74\\
$\log_{10} n_0$ & 2.44 & 0.14 & 0.06 & 0.51 & 0.24\\
$\log_{10} \theta_c$ & 2.31 & 0.07 & 0.02 & 1.04 & 0.87\\
$\log_{10} \epsilon_c$ & 2.74 & 0.04 & 0.03 & 0.68 & 0.48\\
$\log_{10} \epsilon_b$ & 0.62 & 0.18 & 0.07 & 1.64 & 1.38\\
$p$ & 7.49 & 0.03 & 0.41 & 2.77 & 3.58\\
$\log_{10} \xi_N$ & 2.26 & 0.04 & 0.07 & 1.41 & 1.16\\
\end{tabular}
\end{ruledtabular}
\label{tab:eventsKL}
\end{table}

\subsection{Binary progenitor properties and implications for the observed offset of GRB 230307A}\label{offset}

The results presented in the previous subsections suggest that the optical emission observed from GRB 230307A is most consistent with a kilonova scenario originating from the merger of a compact binary system. Within this framework, we now explore the physical properties of the potential progenitor system. Assuming that the kilonova originated from a BNS merger and following \cite{Nicholl2021}, we adopt phenomenological fits from numerical-relativity simulations that relate the ejecta parameters to the binary properties, such as the individual masses and compactnesses. For the dynamical ejecta mass, we use the fitting relations derived by \cite{Dietrich2017}, while the disk mass fits are taken from \cite{Coughlin2019}. We fix the Tolman–Oppenheimer–Volkoff maximum mass to 2.17 M$_{\odot}$ following the constraint reported in \cite{Nicholl2021}. To estimate the neutron star radius, we use the empirical fit for $R_{1.4}$ (the radius of a 1.4 M$_{\odot}$ neutron star) as a function of binary tidal deformability and chirp mass, based on the relations provided in \cite{De2018} and calibrated across different EOS.

\begin{figure}
\centering
\includegraphics[width=\linewidth]{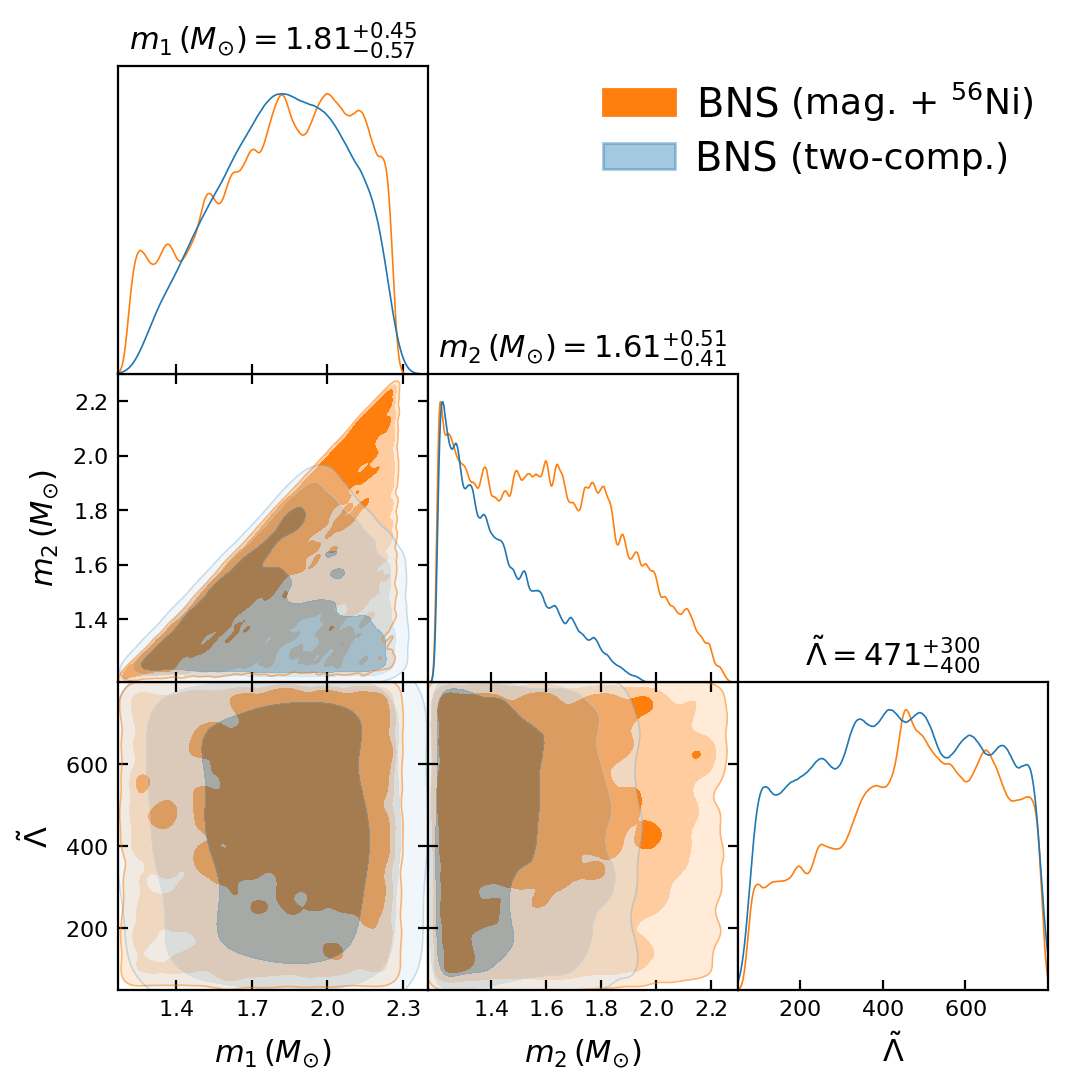}
\caption{Corner plot of the binary component masses ($m_1, m_2$) and dimensionless tidal deformability ($\Lambda$) from inference results of the magnetar spin‑down model (orange) and BNS two-component model (blue). The top panel shows the final inference for the magnetar spin‑down model.}
\label{fig:corner_mass}
\end{figure}

Adopting the model that best fits the observed data, which combines magnetar spin‑down and the radioactive decay of $^{56}$Ni, we infer that the binary system is composed of a primary with a mass of $m_{1} = 1.81^{+0.46}_{-0.61}\, \rm{M}_\odot$ and a secondary with a mass of $m_{2} = 1.61^{+0.65}_{-0.41}\, \rm{M}_{\odot}$, with dimensionless tidal deformability of the binary being $\tilde{\Lambda} = 471^{+318}_{-395}$. Assuming a BNS two-component model for the kilonova emission, we obtain constraints on the binary masses of $m_{1} = 1.82^{+0.45}_{-0.61}\, \rm{M}_\odot$ and $m_{2} = 1.42^{+0.51}_{-0.22}\, \rm{M}_\odot$, together with an associated dimensionless tidal deformability of $\tilde{\Lambda} = 439^{+368}_{-350}$. Although these results suggest a slightly greater asymmetry between the mass components, they remain in overall very good agreement within the uncertainties. These values correspond to the posterior mean estimates with their associated 68\% confidence intervals. For both inferences, we assume uniform prior bounds of [1.2, 2.27] for $m_1$, [1.2, $m_1$] for $m_2$, and [70, 790] for $\tilde{\Lambda}$. The contour plots of these results are shown in Fig.~\ref{fig:corner_mass}.

Based on the inferred BNS component masses, $m_1$ and $m_2$, we search for the conditions under which the binary can attain the projected offset $\ell_\mathrm{obs} \sim 38.9~\mathrm{kpc}$ at coalescence. 
In particular, we consider the scenario in which the binary was orbiting its host galaxy with velocity $\bar{v}_\mathrm{c}$ and suddenly experienced a kick of systemic velocity $\bar{v}'_\mathrm{k}$, where the prime indicates that the quantity is measured in the pre-kick co-moving frame. 
This event leaves the binary with a velocity $\bar{v}_0 = \bar{v}_\mathrm{c} + \bar{v}'_\mathrm{k}$ relative to the galaxy centre. We then investigate the minimal kick velocity $v'_\mathrm{k}$ magnitude required to reproduce the observed kilonova offset. Hereafter, unbarred variables like $v_x$, denote the magnitude of the velocity vector $\bar{v}_x$.

Considering the binary has velocity $v_0$ at a radial location $r_0$ relative to the host-galaxy centre, the classical work–energy theorem gives the binary velocity $v_\mathrm{b}$ at a radial position $r$ as
\begin{equation}
v_\mathrm{b}(r) = \sqrt{v_0^2 + 2[\Phi(r_0) - \Phi(r)]},
\label{v_potential_rel}
\end{equation}
where $\Phi_\mathrm{g}(r) = \Phi_*(r) + \Phi_\mathrm{DM}(r)$ is the total gravitational potential of the galaxy, composed of stellar and dark-matter components, both approached as spherically symmetric, which we define as a function of the stellar mass $M_*$ and redshift $z$, as detailed in App.~\ref{app:grav_potential}.
To estimate the flight time elapsed between ejection and coalescence, $\Delta t_\mathrm{P}$, we assume that the binary velocity is approximately aligned with the radial direction, 
$v_\mathrm{b} \approx dr/dt$, which is valid when the radial distance at coalescence is much larger than the initial radius $r_0$.

Integrating equation~\ref{v_potential_rel} then gives
\begin{align}
\nonumber
\Delta t_\mathrm{P} =&
\frac{5c^5(1+q)^2}{256 G^3 q(m_1+m_2)^3}\frac{a_0^4}{f_e}\\ 
=& 
\int_{r_0}^{\ell_\mathrm{obs}/\cos\theta_\mathrm{p}}  
\frac{dr}{\sqrt{v_0^2+2[\Phi_\mathrm{g}(r_0)-\Phi_\mathrm{g}(r)]} }.
\label{integralAp}
\end{align}
The first equality in equation~\ref{integralAp} corresponds to the coalescence time of a binary with initial separation $a_0$, as derived by \citet{Peters_1964}, where $q=m_2/m_1<1$, and $f_e\sim1$ is a function of the initial binary eccentricity, \textit{G} is the gravitational constant, and \textit{c} is the speed of light. In the second equality, $\ell_\mathrm{obs}/\cos\theta_\mathrm{p}$ is the radial position at coalescence, with $\theta_\mathrm{p}$ the angle between the coalescence direction and the plane of the sky. From equation~\ref{integralAp}, we note that the minimum velocity $v_0$ required for the binary to reach the distance $\ell_\mathrm{obs}/\cos\theta_\mathrm{p}$ is
$v_\mathrm{0,th} = \sqrt{2[\Phi_\mathrm{g}(\ell_\mathrm{obs}/\cos\theta_\mathrm{p}) - \Phi_\mathrm{g}(r_0)]}$.
This corresponds to the threshold value of $v_0$ that ensures the integrand in equation~\ref{integralAp} remains real.

Finally, since $\bar{v}_0 = \bar{v}_\mathrm{c} +\bar{v}'_\mathrm{k}$, the required systemic kick velocity can be obtained as
\begin{equation}
v'_\mathrm{k} = 
\sqrt{v_0^2 - \sin^2\theta_\mathrm{k}v^2_\mathrm{c}}
-v_\mathrm{c}\cos\theta_\mathrm{k},
\label{vk}
\end{equation}
with $\theta_\mathrm{k}$ being the angle among $\bar{v}_\mathrm{c}$ and $\bar{v}'_\mathrm{k}$, and we approach $v_\mathrm{c}$ as the circular velocity of the galaxy at the radius $r_0$ (see App.~\ref{app:grav_potential}, for details).

\begin{figure}
    \centering
    \includegraphics[width=\linewidth]{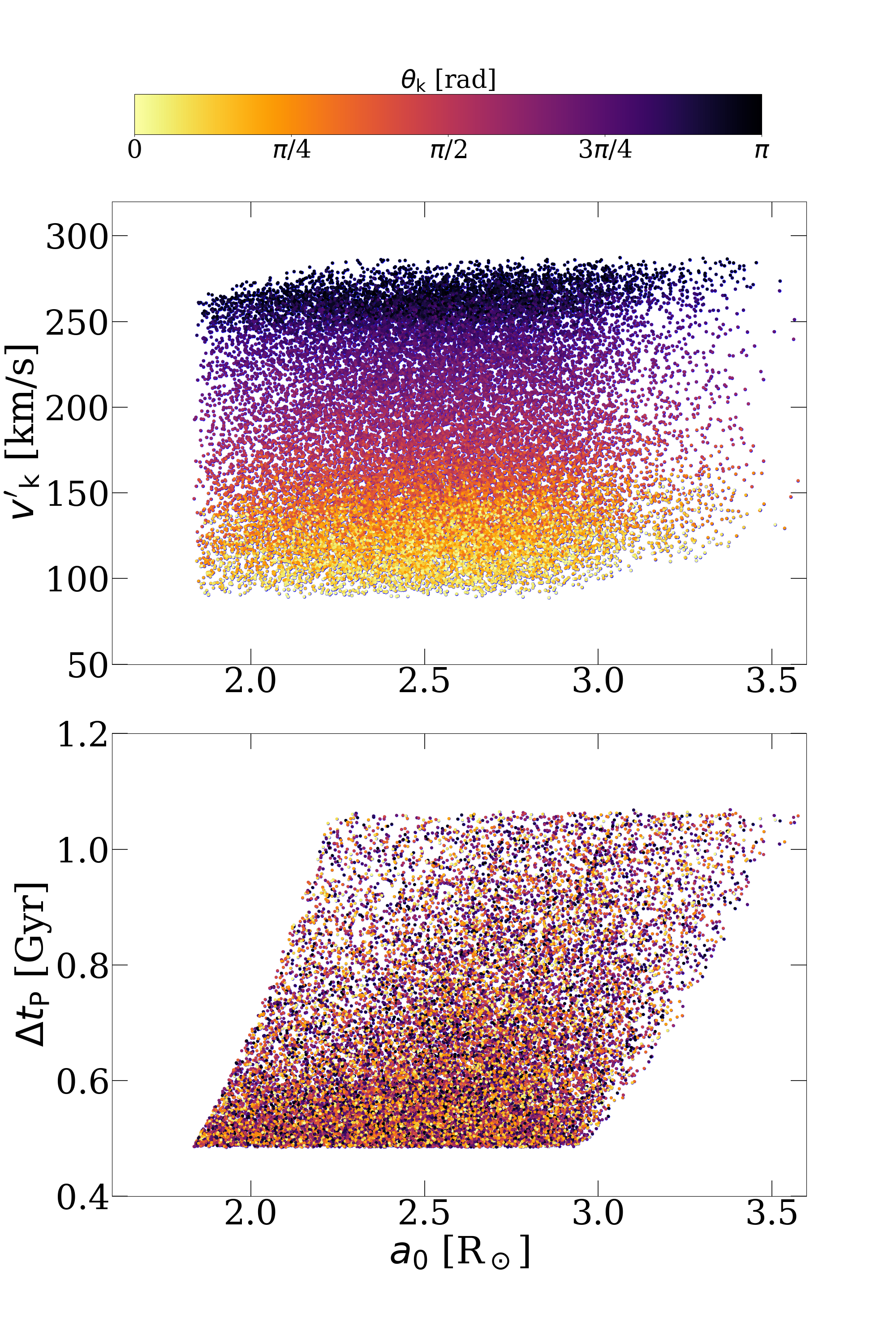}
    \caption{
    Scatter plots for parameter combinations of $v'_\mathrm{k}-a_0$ (upper) and $\Delta t_\mathrm{P}-a_0$ (lower), 
    consistent with the observed merger offset
    $\ell_\mathrm{obs} = 38.9$ kpc, as derived from equations (\ref{integralAp}-\ref{vk}). Colors indicate different assumed values of $\theta_{\mathrm{k}}$.}
    \label{fig:sampled_kick}
\end{figure}

In the upper panel of Fig.~\ref{fig:sampled_kick}, we display the minimum values for $v'_\mathrm{k}$, required to produce the observed offset $\ell_\mathrm{obs} = 38.9$ kpc, as a function of $a_0$ and the angle $\theta_{\mathrm{k}}$.
The points in this $v'_\mathrm{k}-a_0$ space were obtained
by solving the simultaneous equations
(\ref{integralAp}-\ref{vk}), while
fixing the value of $M_*\sim2.4\times10^9$ M$_\odot$ and $z=0.0647$, consistent with the estimate by \cite{Yang2024}, using the posterior samples of $m_1$ and $m_2$ derived from the magnetar spin-down model (see Fig.~\ref{fig:corner_mass}), and uniformly sampling the parameters $\theta_p\in(0,\pi/3)$,  $\theta_\mathrm{k}\in(0,\pi)$, and $r_0\in(0.5,5)$ kpc. In the lower panel of Fig.~\ref{fig:sampled_kick}, we show the coalescence time associated with the parameter configurations of the upper panel.

We observe from Fig.~\ref{fig:sampled_kick} that
the minimum systemic velocity $v'_\mathrm{k}$ required to produce the offset $\ell_\mathrm{obs}=39.8$ kpc
ranges between $\sim100$ and $270~\mathrm{km\,s^{-1}}$.
This velocity clearly depends on the
kick angle $\theta_\mathrm{k}$, with smaller values of $\theta_\mathrm{k}$ leading to the lowest $v'_\mathrm{k}$.
The corresponding binary separations $a_0$ mostly lie within
$2$–$3~R_\odot$, regardless of the kick direction. Under such conditions, the ejections would produce the electromagnetic transient within $\Delta t_\mathrm{P}\lesssim 1$ Gyr, as shown in the lower panel of Fig.~\ref{fig:sampled_kick}.
We note that this period of time is
comparable to the galaxy evolution time scale, and much lower than the Hubble time (in $\Lambda$CDM cosmological models).

A possible explanation for the origin of the sudden ejection of the binary is the natal kick imparted to one of its components during an asymmetric supernova explosion \cite{Brandt_1995}. 
This scenario is particularly plausible for systemic kicks that are quasi-aligned with the binary’s orbital motion before the supernova event (i.e. $\theta_\mathrm{k} \lesssim \pi/4$), 
which would imply systemic velocities of 
$v'_\mathrm{k} \sim 100$–$150~\mathrm{km\,s^{-1}}$, 
as shown in Fig.~\ref{fig:sampled_kick}. 
Natal kicks producing such parameter configurations of $v'_\mathrm{k}$ and $a_0$ would require finely tuned conditions to avoid binary disruption, yet remain possible according to analyses of similar phenomena discussed in \cite{Tauris2017, Renzo2019} and are consistent with the inferred natal kick velocities of some observed Galactic BNS pulsars \cite{Disberg2024}.

\section{Conclusions}\label{conclusion}
GRB 230307A is the second-brightest gamma-ray burst ever observed. The absence of an associated supernova, combined with persistent late-time near-infrared emission and rapidly decaying optical emission, along with spectroscopic detection of emission lines indicative of heavy elements such as tellurium, strongly supports the presence of a kilonova powered by $r$-process nucleosynthesis. The long duration of GRB 230307A initially suggested a traditional classification as the result of the collapse of a massive star. However, the presence of soft X-ray emission characterized by a plateau followed by a power-law decay points to the presence of a magnetar central engine, which is typically associated with CBC. Furthermore, the large observed offset from the host galaxy provides additional support for a CBC origin. These factors pose a challenge to establishing a direct and comprehensive understanding of the progenitor of GRB 230307A. In Section \ref{modelselection}, we present a detailed Bayesian multi-wavelength analysis of GRB 230307A, examining various progenitor scenarios such as BNS, NS–WD, NS–BH mergers, and a white dwarf TDE. Model comparison based on the Bayes factor and LOO score strongly favors a BNS or NS–WD progenitor, in which the emission is powered by magnetar spin-down and ${}^{56}\mathrm{Ni}$ radioactive decay. This conclusion aligns with previous works \cite{Du2024, Sun2025}, which analyzed prompt X-ray and gamma-ray data from LEIA and GECAM and reached similar findings. Our analysis further demonstrates that a kilonova component, in addition to the standard afterglow emission, is essential to reproduce the observations, underscoring the complex nature of the emission mechanisms in GRB 230307A.

We also investigated the source properties of the possible progenitor of GRB 230307A. Assuming a BNS progenitor, we used the relations between the ejecta properties and the masses and tidal deformabilities of the binary components as described by \citet{Dietrich2017} and \citet{Coughlin2019}. Applying these relations to the two best-fit models, the magnetar spin-down and the two-component kilonova, we performed a Bayesian inference to constrain the BNS source properties. Both models yielded consistent results, indicating component masses of $m_{1} = 1.81^{+0.46}_{-0.61}\,\rm{M}_{\odot}$ and $m_{2} = 1.61^{+0.65}_{-0.41}\,\rm{M}_{\odot}$, with a tidal deformability of $\tilde{\Lambda} = 471^{+318}_{-395}$ for the magnetar spin-down model. Moreover, we used the posterior mass samples to investigate the ejection conditions of the binary progenitor responsible for the large projected merger offset. 
Assuming the system was ejected from the nearest galaxy by a kick imparted on top of its orbital motion, we find that the minimum systemic velocity $v'_\mathrm{k}$ (in the co-moving frame) required to reproduce the observed offset of $\ell_\mathrm{obs}=39.8$~kpc ranges between $\sim100$ and $270~\mathrm{km\,s^{-1}}$. 
The corresponding post-kick binary separation falls within $a_0\sim2$–$3~R_\odot$. 
If the kick is quasi-aligned with the pre-kick orbital motion, the required velocity decreases to $v_\mathrm{k}\sim100$–$150~\mathrm{km\,s^{-1}}$. 
This special case can be plausibly explained within the natal-kick scenario, where one binary component undergoes an asymmetric supernova explosion, under finely tuned conditions that prevent binary disruption \cite{Tauris2017, Renzo2019}.

The results of this work provide statistical evidence supporting a CBC origin for GRB 230307A. The inferred parameter values from the magnetar spin-down model are consistent with either a BNS merger or a NS-WD merger. Distinguishing between these two progenitor channels remains challenging, primarily due to the limited availability of detailed numerical simulations for NS-WD mergers, in contrast to the extensive modeling developed for BNS systems (e.g., \cite{PhysRevLett.131.011401, Murguia-Berthier:2021tnt, Combi:2022nhg, rezzolla2013relativistic, Coughlin:2018fis, Dietrich:2016fpt, Radice:2018pdn, Nedora:2020qtd}). Without comparable simulation data, including detailed predictions of electromagnetic counterparts and ejecta properties, it is currently not possible to determine the nature of the progenitor based solely on observational features.

Future progress will require the development of a systematic suite of general-relativistic hydrodynamic simulations targeting mergers between neutron stars and white dwarfs. This poses unique technical challenges due to the broad range of spatial and temporal scales required to accurately model both types of stellar remnants with realistic equations of state. These simulations must also incorporate variations in mass ratios, white dwarf compositions, and neutron star properties \cite{Moran-Fraile:2023oui}. Such efforts are essential for producing reliable predictions of observables, including kilonova light curves, nucleosynthetic yields, and gravitational wave signatures specific to these systems. Ultimately, this work would provide the necessary foundation for distinguishing between different types of compact binary progenitors in future multi-messenger observations.

Gravitational waves emitted during the coalescence could provide complementary observational constraints to resolve this ambiguity. Unfortunately, the distance of GRB 230307A is slightly beyond the current sensitivity of the LIGO, Virgo, and KAGRA detectors, meaning that even if the event had occurred during the fourth observing run, it would have been undetectable. Future GW detectors, with sensitivity extending beyond the distance of GRB 230307A, will open a promising opportunity for multi-messenger astronomy, providing complementary insights into its progenitors.

\section*{Acknowledgements}
The authors made use of Sci-Mind servers machines developed by the CBPF AI LAB team and would like to thank Paulo Russano and Marcelo Portes de Albuquerque for all the support in infrastructure matters. J.C.R.R. acknowledges support from Rio de Janeiro State Funding Agency FAPERJ, grant E-26/205.635/2022.
C.R.B. acknowledges the financial support from CNPq (316072/2021-4) and from FAPERJ (grants 201.456/2022 and 210.330/2022) and the FINEP contract 01.22.0505.00 (ref. 1891/22). 

\appendix
\section{Parameter inference from multi-wavelength observations}\label{app:inference_resu}
In this appendix, we present the results of a joint Bayesian inference from the multi-wavelength analysis of GRB 230307A, assuming that the emission consists of an afterglow signature and a transient emission (either kilonova or TDE). The Bayesian inference was performed using a dynamic nested sampling algorithm provided by the \textsc{dynesty} Python package. In each simulation, we employed 500 live points along with a multi-ellipsoidal bounding method, and terminated the sampling process when the change in log-evidence dropped below a tolerance threshold of 0.1. The median values and the 1$\sigma$ confidence intervals are presented in Table \ref{tab:bestfit}.

\begin{table*}
\caption{Posterior medians and 1$\sigma$ credible intervals for the model parameters. The GRB afterglow model assumes a Gaussian jet structure with the luminosity distance fixed at 291 Mpc. Prior bounds are indicated in brackets in the first column; all priors are assumed to be uniform.}
\begin{ruledtabular}
\begin{tabular}{cccccc}%{llllll}
 & CBC (mag.+${}^{56}\mathrm{Ni}$) & BNS (two-comp.) & BNS (cental eng.) & NS-BH & TDE \\
\hline\hline
GRB model parameter \\
estimation - [prior] \\
\hline 
$\log_{10}E_{\rm{K, iso}}$ (erg) - [50, 60] & 53.18$^{+1.74}_{-1.58}$ & 53.11$^{+1.18}_{-1.24}$ & 53.36$^{+1.59}_{-1.52}$ & 53.86$^{+0.56}_{-1.03}$ & 53.34$^{+0.49}_{-0.48}$ \\
$\log_{10}n_{0}$ (cm$^{-3}$) - [-6, 2] & -0.75$^{+1.99}_{-2.32}$ & -0.53$^{+1.90}_{-4.91}$ & -0.30$^{+1.80}_{-2.41}$ & 0.05$^{+0.97}_{-0.27}$ & 0.20$^{+1.61}_{-1.57}$\\
$\log_{10}\theta_{\rm c}$ - [-2, -0.5] & -0.89$^{+0.38}_{-0.55}$ & -0.56$^{+0.06}_{-0.94}$ & -0.88$^{+0.36}_{-0.54}$ & -1.26$^{+0.22}_{-0.10}$ & -1.23$^{+0.10}_{-0.20}$\\
$\log_{10}\epsilon_{e}$ - [-6, -0.3] & -2.68$^{+1.40}_{-1.46}$ & -2.41$^{+2.10}_{-3.55}$ & -2.96$^{+1.48}_{-1.62}$ & -3.53$^{+1.07}_{-0.65}$ &  -2.84$^{+0.57}_{-0.46}$\\
$\log_{10}\epsilon_{B}$ - [-6, -0.3] & -3.22$^{+1.31}_{-1.22}$ & -2.61$^{+2.31}_{-2.37}$ & -3.56$^{+1.25}_{-1.05}$ & -1.00$^{+0.67}_{-0.47}$ & -1.63$^{+0.87}_{-0.59}$\\
$p$ - [2.01, 2.9] & 2.12$^{+0.09}_{-0.10}$ & 2.14$^{+0.18}_{-0.12}$ & 2.13$^{+0.06}_{-0.09}$ & 2.80$^{+0.10}_{-0.11}$ & 2.81$^{+0.09}_{-0.12}$ \\
$\log_{10}\xi_{N}$ - [-5, 0] & -2.36$^{+1.23}_{-1.13}$ & -2.37$^{+1.30}_{-1.42}$ & -2.75$^{+1.40}_{-1.32}$ & -4.31$^{+0.85}_{-0.63}$ & -3.95$^{+0.66}_{-0.65}$\\
\hline\hline
KN model parameter estimation \\
- CBC (mag.+${}^{56}\mathrm{Ni}$)\\
\hline 
$\log_{10}\kappa$ (cm$^2$g$^{-1}$) - [-1, 0] & $-0.34^{+0.24}_{-0.25}$ & \ldots & \ldots & \ldots & \ldots \\
$\log_{10}M_{\rm{ej}}$ (M$_{\odot}$) - [-2.0, -0.8] & $-1.22^{+0.29}_{-0.32}$ &  \ldots & \ldots & \ldots & \ldots \\
$\log_{10}M_{\rm{Ni}}$ (M$_{\odot}$) - [-5.0, -3.0] & $-3.40^{+0.10}_{-0.02}$ & \ldots & \ldots & \ldots & \ldots \\
$\log_{10}\chi\rm{L}_{\rm{sd}}\left(0\right)$ (erg s$^{-1}$) - [45.0, 48.5] & $47.66^{+0.37}_{-0.52}$ & \ldots & \ldots & \ldots & \ldots \\
$v_{\rm min}$ (c) - [0.001, 0.15] & $0.10^{+0.01}_{-0.01}$ & \ldots & \ldots & \ldots & \ldots \\
$v_{\rm max}$ (c) - [0.18, 0.35] & $0.26^{+0.04}_{-0.04}$ & \ldots & \ldots & \ldots & \ldots \\
$\delta$ - [1.0, 3.0] & $1.74^{+0.70}_{-0.62}$ & \ldots & \ldots & \ldots & \ldots \\
\hline\hline
KN model parameter estimation \\
- BNS (two-comp.)\\
\hline 
$\log_{10}M_{\rm{ej,1}}$ (M$_{\odot}$) - [-3, -1] & \ldots & $-1.18^{+0.18}_{-0.08}$ & \ldots & \ldots & \ldots \\
$\log_{10}v_{\rm{ej,1}}$ (c) - [-1.0, -0.5] & \ldots &  $-0.90^{+0.05}_{-0.07}$ & \ldots & \ldots & \ldots \\
$\log_{10}\kappa_{1}$ (cm$^{2}$ g$^{-1}$) - [-2.0, 0.5] & \ldots & $-0.41^{+0.42}_{-0.40}$& \ldots & \ldots & \ldots \\
$\beta_{v,1}$ - [1, 5] & \ldots & $2.47^{+1.39}_{-1.27}$ & \ldots & \ldots & \ldots \\
$Y_{e,1}$ - [0.2, 0.4] & \ldots & $0.31^{+0.06}_{-0.07}$ & \ldots & \ldots & \ldots \\
$\log_{10}M_{\rm{ej,2}}$ (M$_{\odot}$) - [-3, -1] & \ldots & $-1.11^{+0.10}_{-0.02}$& \ldots & \ldots & \ldots \\
$\log_{10}v_{\rm{ej,2}}$ (c) - [-2, -1] & \ldots & $-1.34^{+0.11}_{-0.10}$& \ldots & \ldots & \ldots \\
$\log_{10}\kappa_{2}$ (cm$^{2}$ g$^{-1}$) - [-0.5, 2.0] & \ldots & $1.50^{+0.39}_{-0.35}$ & \ldots & \ldots & \ldots \\
$\beta_{v,2}$ - [1, 5] & \ldots & $3.34^{+1.26}_{-1.14}$ & \ldots & \ldots & \ldots \\
$Y_{e,2}$ - [0.1, 0.2] & \ldots & $0.14^{+0.03}_{-0.03}$ & \ldots & \ldots & \ldots \\
\hline\hline
KN model parameter estimation \\
- BNS (central eng.)\\
\hline 
$\log_{10}M_{\rm{ej}}$ (M$_{\odot}$) - [-3, -1] & \ldots &  \ldots & $-1.84^{+0.12}_{-0.11}$ & \ldots & \ldots \\
$\log_{10}v_{\rm{ej}}$ (c) - [-2, -0.5] & \ldots &  \ldots & $-1.50^{+0.20}_{-0.31}$ & \ldots & \ldots \\
$\zeta$ - [0, 0.5] & \ldots & \ldots & $0.20^{+0.16}_{-0.15}$ & \ldots & \ldots \\
$\kappa$ (cm$^2$g$^{-1}$) - [0.1, 10.0] & \ldots & \ldots & $3.01^{+1.56}_{-1.61}$ & \ldots & \ldots \\
$\log_{10}\rm{L}_{0}$ (erg s$^{-1}$) - [45.5, 50.0] & \ldots & \ldots & $48.06^{+0.63}_{-0.69}$ & \ldots & \ldots \\
$n_{\rm{ism}}$ (cm$^{-3}$) - [3.0, 5.0] & \ldots & \ldots & $3.47^{+0.60}_{-0.42}$ & \ldots & \ldots \\
\hline\hline
KN model parameter estimation \\
NS-BH - [prior]\\
\hline 
$M_{\rm{BH}}$ (M$_{\odot}$) - [5.0, 100.0] & \ldots &  \ldots & \ldots & $49.99^{+23.73}_{-17.33}$ & \ldots \\
$M_{\rm{NS}}$ (M$_{\odot}$) - [1.2, 2.27] & \ldots &  \ldots & \ldots & $1.49^{+0.33}_{-0.24}$ & \ldots \\
$\chi_{\rm{BH}}$ - [0.05, 1.0] & \ldots & \ldots & \ldots & $0.68^{+0.20}_{-0.26}$ & \ldots \\
$\Lambda_{\rm{NS}}$ - [5.0, 5000.0] & \ldots & \ldots & \ldots & $1396^{+1548}_{-1163}$ & \ldots \\
$\zeta$ - [0, 0.5] & \ldots & \ldots & \ldots & $0.33^{+0.13}_{-0.17}$ & \ldots \\
$\log_{10}v_{\rm{ej,2}}$ (c) - [-2.0, -0.5] & \ldots & \ldots & \ldots & $-1.48^{+0.59}_{-0.41}$ & \ldots \\
$\log_{10}\kappa_{1}$ (cm$^2$g$^{-1}$) - [-2.0, 0.5] & \ldots & \ldots & \ldots & $-1.48^{+0.59}_{-0.41}$ & \ldots \\
$\log_{10}\kappa_{2}$ (cm$^2$g$^{-1}$) - [-0.5, 2.0] & \ldots & \ldots & \ldots & $-0.46^{+0.53}_{-0.59}$ & \ldots \\
\hline\hline
KN model parameter estimation \\
TDE - [prior]\\
\hline 
$M_{\rm{BH}}$ ($10^6$ M$_{\odot}$) - [0.1, 100.0] & \ldots & \ldots & \ldots & \ldots & $0.45^{+1.39}_{-0.34}$ \\
$M_{\rm{star}}$ (M$_{\odot}$) - [0, 1.44] & \ldots &  \ldots & \ldots & \ldots & $0.69^{+0.43}_{-0.49}$ \\
$t_{\rm{visc.}}$ (days) - [1e$^{-3}$, 1e$^5$] & \ldots & \ldots & \ldots & \ldots & $72669^{+18785}_{-16641}$ \\
$b$ - [0,2] & \ldots & \ldots & \ldots & \ldots & $0.72^{+0.77}_{-0.64}$ \\
$\eta$ - [0.005,0.4] & \ldots & \ldots & \ldots & \ldots & $0.13^{+0.13}_{-0.11}$ \\
$\log_{10} L_{\rm{Edd}}$ (erg s$^{-1}$) - [43.1, 46.1]& \ldots & \ldots & \ldots & \ldots & $43.66^{+0.57}_{-0.48}$ \\
$\log_{10} R_{\rm photo}$ (km) - [-4,4] & \ldots & \ldots & \ldots & \ldots & $-1.69^{+2.06}_{-1.60}$ \\
$\rm{L}_{\rm photo}$ (erg s$^{-1}$) - [0,4] & \ldots & \ldots & \ldots & \ldots & $2.77^{+1.12}_{-1.17}$ \\
\end{tabular}
\end{ruledtabular}
\label{tab:bestfit}
\end{table*}

\section{The gravitational potential of the host galaxy}\label{app:grav_potential}
We model the gravitational potential, $\Phi_\mathrm{g}$, of the putative host galaxy that formed the progenitor binary system discussed in the main text as spherically symmetric, and composed of stellar and dark matter (DM) components:
\begin{equation}
\Phi_\mathrm{g}\left(r\right) = \Phi_*\left(r\right) + \Phi_\mathrm{DM}\left(r\right).
\label{Phi_g}
\end{equation}

For the stellar component, we adopt the Hernquist profile \cite{Hernquist_1990}:
\begin{equation}
\label{Pot_hq}
\Phi_*\left(r\right) = - \frac{G M_*}{r + a_\mathrm{h}},
\end{equation}
where $M_*$ is the total stellar mass, and $a_\mathrm{h}$ is the scale radius, assumed to be $a_\mathrm{h} = 0.55\,R_{50}$ \cite{Abbott_2017}, with $R_{50}$ the stellar half-mass radius. The latter is obtained as a function of $M_*$ from the empirical relation derived in \cite{Zhang2019}.

For the DM component, we adopt the Navarro–Frenk–White (NFW) potential \cite{Navarro_1996}:
\begin{equation}
\Phi_\mathrm{DM}\left(r\right) = -\frac{4\pi G \rho_0 R_\mathrm{s}^3}{r}
\ln\left(1 + \frac{r}{R_\mathrm{s}}\right),
\end{equation}
where $\rho_0$ is the characteristic density of the DM halo, and $R_\mathrm{s}$ its scale radius.
These parameters are related to the halo mass, $M_\mathrm{DM}$, through
\begin{equation}
M_\mathrm{DM} = 4\pi \rho_0 R_\mathrm{s}^3 \left[
\ln\left(1+\mathbb{c}\right) - \frac{\mathbb{c}}{1+\mathbb{c}}
\right],
\label{MDM}
\end{equation}
where $\mathbb{c}$ is the concentration parameter.

We estimate $M_\mathrm{DM}$ as a function of the total stellar mass $M_*$ and the source redshift, using the empirical relation derived by \cite{Girelli2020}. The concentration parameter $c$ is computed using the \texttt{colossus} toolkit \cite{Diemer_2018}, specifying $M_{200}$, the source redshift, and the mass–concentration relation from \citet{Ishiyama2021}.  
Given $\mathbb{c}$ and $M_\mathrm{DM}$, the scale radius is defined as $R_\mathrm{s} = R_\mathrm{vir}/\mathbb{c}$, where
\begin{equation}
R_\mathrm{vir} = \left[\frac{3 M_\mathrm{DM}}{4\pi \Delta_\mathrm{vir} \rho_\mathrm{c}}\right]^{1/3},
\end{equation}
with $\Delta_\mathrm{vir} = 200$ and the critical density of the Universe, $\rho_\mathrm{c} = 1.4 \times 10^{11}\,M_\odot\,\mathrm{Mpc^{-3}}$.
Finally, the characteristic density $\rho_0$ is obtained from equation~\ref{MDM}.

With the definitions above, the galactic potential in equation~\ref{Phi_g} is fully determined by specifying the total stellar mass $M_*$, the source redshift $z$, and the radial position $r$. 
Finally, given the gravitational potentials, the implied circular velocity of the galaxy at radius $r$ is
\begin{equation}
v_\mathrm{c}\left(r\right) = \sqrt{r \left( \frac{d\Phi_*}{dr} + \frac{d\Phi_\mathrm{DM}}{dr} \right)}.
\end{equation}

\bibliography{apssamp}% Produces the bibliography via BibTeX.

\end{document}